# Exploring Practitioner Perspectives of Sourcing Risks: Towards the Development of an Integrated Risk and Control Framework

**Deborah Bunker**
The University of Sydney Business School
University of Sydney
Sydney, New South Wales
Email: deborah.bunker@sydney.edu.au

**Catherine Hardy**
The University of Sydney Business School
University of Sydney
Sydney, New South Wales
Email: catherine.hardy@sydney.edu.au

**Abdul Babar**
The University of Sydney Business School
University of Sydney
Sydney, New South Wales
Email: abdul.babar@sydney.edu.au

**Kenneth J Stevens**
UNSW Business School
University of New South Wales
Sydney, New South Wales
Email: k.stevens@unsw.edu.au

## Abstract

Outsourcing of information and communication technologies (ICT) and related services is an established and growing industry. Recent trends, such as the move toward multi-sourcing have increased the complexity and risk of these outsourcing arrangements. There is a critical research need to identify the risks faced by both the organisations that outsource ICT and the vendors that provide it in this changing landscape. To address growing concerns regarding the best way to deal with risk and control in this environment, our research focuses on establishing a Sourcing Risk and Control Framework to assist organisations identify these risks and develop effective mitigation strategies. In this paper we report on the first stage of our research that sought to document how sourcing risk is represented and considered in practice. To date, limited empirical research has been conducted in an Australian context. Using a series of workshops involving client and vendor representatives, we identified a broad range of risks and developed a cohesive categorisation scheme that incorporates functional and multi-stakeholder perspectives.

## 1    Introduction - The Business and Research Imperative

The management of sourcing risks, and their various guises, has been a focal area of research in the information systems (IS) field for close to two decades (eg. Earl 1996; Willcocks et al. 1999; Warkentin and Adams 2007; Herath and Kishore 2009; Yim 2014). Gonzalez et al. (2013) argue that within the outsourcing domain, risk is second only to success in terms of importance. While sourcing risk has been extensively researched, it remains a theoretical and practical challenge, especially given the variation in risks due to differences in the scope of services sourced (Jain and Thietart 2013) and the increasing challenge of integrating services across multiple service providers (Deloitte 2013a; ISACA 2014). Recent technological developments such as cloud computing, and broadband and mobile technologies have revealed a dynamic and inter-dependent nature of risks, particularly in operational security risks (Rocco Grillo cited in Protiviti 2014), that is yet to be fully understood.

As the "scope, scale and complexity of vendor relationships and services increase," (ISACA 2014, p.9), the effective identification of sourcing risks, their management and implementation of cost effective controls is paramount to organisational productivity, performance, growth and sustainability. Yet, recent industry surveys reveal that organisations are failing to adequately address these risks and lack the necessary capabilities to manage them effectively (Protiviti 2014). In addition, as global sourcing expands, so do the types of sourcing models and standards/regulations that guide these practices (e.g. Sandeep et al. 2013). These issues focus attention on the need for developing new approaches and a comprehensive risk management strategy to effectively manage these risks at a practical (e.g. Deloitte 2013b; ISACA 2014; Protiviti 2014) and theoretical level (e.g. Lacity et al. 2010; Mathew 2011).

In response to these challenges we have developed a major research project with the aim of developing an innovative and integrated framework that can support business to effectively conduct sourcing risk assessments and implement appropriate mitigation strategies. Concurrently, the research team is also developing an ontological representation of this framework with the objective of creating an interactive website to facilitate and incorporate ongoing input to the framework on both a local and global scale.

Overall our project has three objectives: 1) understanding the implications for the design and use of such a framework in organisations; 2) making the framework part of current sourcing risk identification and management practices of an organisation; and 3) developing an understanding of sourcing risk in an Australian context. We seek to address issues such as: critical uncertainties of sourcing arrangements; the power imbalance between supplier and customer; complex compliance and standards; the nature of the supplier and customer organisation and its effect on sourcing arrangements; the nature of the sourcing transaction; and leverage and negotiation mechanisms.

There are four stages to this research program, which we outline in section two. The aim of this paper is to report on the first piece of work from Stage 1, which is designed to address the following question: "How do organisations identify and classify risks in their sourcing arrangements and collaborations?"

The structure of the paper is as follows. Firstly, we provide background information about the research project. We then follow this discussion with the literature review. The third section outlines the overall project methodology and workshop format for Stage 1. This is followed by the preliminary risk classification arising from the workshops. Next the implications from the first stage of our research are discussed, followed by the conclusion and future directions for our research project.

## 2   Research Background

The overall objective of the project is to develop a new and innovative framework which can be applied by all parties to a sourcing transaction (supplier and customer) to 1) decrease the rate of failure of sourcing arrangements by ensuring that the most cost effective controls are implemented and used, and 2) decrease the transaction costs for sourcing by limiting the use of inappropriate or ineffective controls and by encouraging the selection of controls that are appropriate and effective. When established, the Sourcing Risk and Control Framework could be used for further research into novel control strategies including (but not limited to) incomplete contracts as well as strategies based on enhancing social interactions between suppliers and purchasers.

### 2.1   Project Background

In order to develop a Risk Management and Control Framework, a number of workshops with a cross-section of industry participants were conducted.  The objectives of the workshop were to determine:

- how practitioners identify and manage risk complexity through their patterns of control within their sourcing arrangements; and
- the technical, social and institutional influences embedded in their risk perceptions.

Having developed the framework we would then test and continually improve it through a series of continuous case studies of organisations in different industrial settings as well as through the development of an interactive website to facilitate and incorporate ongoing input to the framework. We would do this in order to produce findings based on the identification of overall patterns or individual approaches to the topic and highlight the implications for ICT sourcing arrangements and collaborations from a risk management perspective.

The project has 4 distinct research stages outlined as follows:

**Stage 1** - Develop sourcing risk identification and classification (workshops);

**Stage 2** – Develop control patterns (i.e. identify enablers, inhibiters and mechanisms to control sourcing risks) (workshops);

**Stage 3** – Integrate sourcing risks and classifications as well as control patterns in a framework (workshops); and

**Stage 4** – Apply and test the framework to identify risks and controls as well as measure control effectiveness to inform decisions and improve the framework within each case organisation.

This paper outlines Stage 1 findings and reports on the workshops held with a cross-section of industry representatives, which were designed to:

- Address the current limited understanding of risks in organisational sourcing arrangements that is impacting upon effective decision making on outsourcing by organisations;

- Develop meta-level learning across organisations, which has the potential to reduce the impact and cost of sourcing risks;

- Encourage academic and industry information sharing about ICT sourcing risk identification, to facilitate learning; and

- Develop a risk classification as a basis for the long-term development of a Sourcing Risk and Control Framework.

Upon the completion of Stage 2 of the project (as outlined above), our framework will also include the development of risk controls that should be applied to sourcing risks, the overall objectives of these controls, as well as the perceived effectiveness of such controls in different sourcing situations.

## 2.2　Literature Review

Failure to effectively manage issues/factors such as the identification of appropriate providers, the identification of clear outsourcing objectives understood by all stakeholders, provider attention to client problems, frequent provider/client contact, value for money arrangements, top management support, and appropriate contract structures, present risks to a sourcing arrangement (Gonzales et al. 2008, Hirschheim 2009, Goo et al. 2009, Lacity et al. 2010, Gonzalez et al 2013). Various risk types and classifications have been proposed by numerous studies (see for eg. Herath and Kishore 2009; Nakatsu and Iacovou 2009; de Sà-Soares et al. 2014) and typically include categories such as client/vendor capabilities, supply risk, strategic, legal/regulatory risks, financial, geopolitical, technology, strategic, environmental and sustainability, reputation, employee morale and process and control risks. Whilst these risk classification studies are useful, they have mostly based on literature reviews. Empirical studies that have been undertaken have been conducted in the USA (eg. Kim and Chai 2014), Europe (see Lacity et al. 2010 review) or Asia (Lam 2011; Qin et al 2012). Few empirical sourcing risk studies have been conducted in Australia for example Cullen et al. 2005, Rouse & Corbitt 2003, and Rouse & Corbitt, 2007 . The importance of understanding contextual factors in analysing sourcing risks was highlighted by Willcocks et al. (1999) and is of increasing significance given the geographical dispersion of sourcing arrangements and possible risk exposures.

The increasing inter-connected nature of sourcing risks requires an inter-disciplinary and multi-stakeholder view to bring together the different perceptions and approaches that can be employed in risk management. Limited attention has been directed towards exploring: the relationship of these risk factors to one another and their prioritisation (Gandhi et al 2012); the type of controls that need to be put in place to mitigate the risk to an organisation's sourcing arrangements (Wullenweber et al. 2008; Mathew 2011); or how to measure the effectiveness of these controls. In addition, there have been calls for more "holistic and rich theoretical perspectives" to be used in the sourcing domain (Fregtag et al. 2012). Theories such as transaction cost theory, agency theory and resource-based theory have been dominant in explaining sourcing motivations and risks (see Appendix A in Gonzalez 2010). Whilst useful, these theories provide limited insight into risk perceptions influenced by cultural, socio-political, and cognitive factors such as past experiences (see for e.g. Gorla and Lau 2010). Perceptions of risks from multiple stakeholder groups, in different industry and national contexts are critical in informing empirical research and may provide insights for practitioners to understand 'each others' perceptions of risk.

## 3　Project Methodology

The project utilizes and applies a proven methodological approach (stakeholder workshops from a Discovery perspective) to develop and implement a Sourcing Risk and Control Framework. This research approach is critical to the effectiveness of the data collection and analysis within this study due to the variable nature of organisational outsourcing requirements, diversity of risks associated with these requirements, the variation in control types that can be applied and the multitude of methods that can be used to assess them. That is to say, more quantitative methods of data collection (such as experiments, surveys and field studies) would not gather the most appropriate data for interpretation so to detect the subtleties and differences between organisations. The key generic components of Discovery and Action Research methodologies that are of particular relevance to the development and implementation of the Sourcing Risk and Control Framework are:

Workshops –The *diagnostic* component, involving researchers and key industry representatives in developing a shared interpretation of the Sourcing Risk and Control Framework objectives, assumptions, information, processes and support practices; diagnosis also involves problems related to implementation of a particular framework design and achievements of the framework objectives;

Workshops – The *intervention* component (also called therapeutic), involving the design and re-design of the Sourcing Risk and Control Framework objectives, assumptions, information, processes and support practices, based on diagnosis; and

Organisational Case Studies – The *learning* component, involving distinct, ongoing processes of reflection on consultative practices underway and learning from observations of changes in these practices in the design of the Sourcing Risk and Control Framework. This will be undertaken in the context of the critical argument theory.

While there are several different models and forms of action research, the most appropriate for this study is the canonical form as it implies a cyclic, reflective, iterative and rigorous process (Baskerville and Wood-Harper 1998). Each cycle in this process involves phases of diagnosing, action planning, action taking, evaluating and specifying learning (Fig 1).

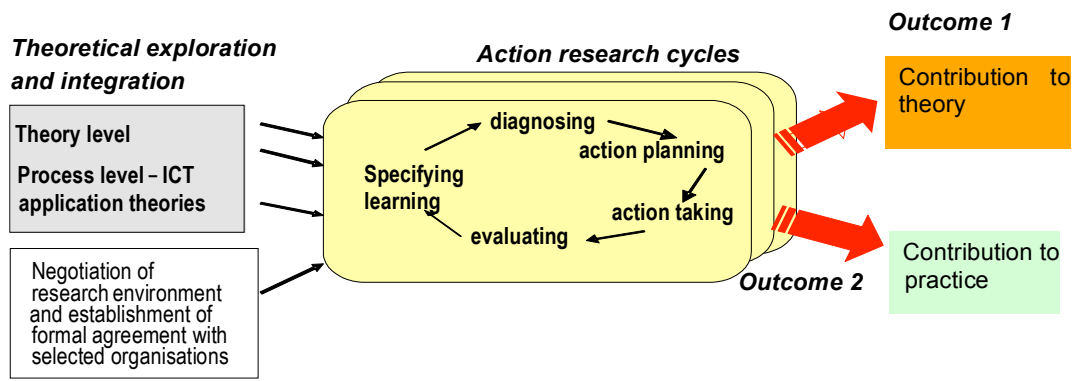

**Figure 1: Research design** - adapted from Bunker & Smith 2009

Data is being collected throughout the Discovery and Action Research cycles by the project team. Formal records of all the meetings with organisers and participants of consultations are being kept; the collaborative reflection and learning phases are being recorded (with the permission of participants); interviews are to be conducted with selected participants in a case consultation; and records of consultations are all being archived for subsequent analysis. This approach has been applied to other industry sectors for solution derivation and implementation as well as development of innovations (Bunker et al. 2007; Pang and Bunker 2005; Smith et al. 2010).

### 3.1  Stage 1 Diagnostic Workshops – Risk Identification and Classification

A number of workshops were held with practitioners in the latter half of 2014 so as to develop an understanding of risks and their relationships. In order to better prepare for the workshop activities, the research team conducted an "expert walk-though" with a skilled outsourcing practitioner where a preliminary list of risks was identified.

Two workshops were subsequently conducted with both practitioners and academics that had expertise in sourcing risk and control. As the opportunity to gather expert practitioners in a workshop setting was highly time-constrained it was decided to use the Quality Function Deployment (QFD) method to facilitate: 1) the development of the risk lists: and 2) their categorisation (e.g. Crow 1994, Akao 2004).

A third case-based role playing workshop was held with a few practitioners and academics to apply the risks and categories that were developed from the first two workshops and to test their rigour and relevance. Workshop participants are detailed in Table 1.

| **Practitioners** | **Count** | **Academics** | **Count (discipline)** |
|---|---|---|---|
| Director | 4 | Professor | 1 (IS) |
| Information security manager | 1 | Senior Lecturer | 3 (IS, computer science) |
| Legal council | 1 | Lecturer | 2 (IS, accounting) |
| Practice lead | 1 | | |
| Project manager | 1 | | |
| Architect/technical lead | 2 | | |
| | 10 | | 6 |

*Table 1. Workshop participants*

The workshops participants were drawn from a number of industries, including consulting, banking and finance, and insurance as well as professional bodies and were selected for their considerable knowledge and experience in IT outsourcing.

| **Timing** | **Activities** | **Outcomes** |
|---|---|---|
| First workshop | <ul><li>Explanation of the QFD method</li><li>Discussion about risk requirements and categories</li><li>Requirements classified into categories</li><li>Risks identified and written on PostIt notes</li></ul> | <ul><li>Framework requirements</li><li>Risk categories identified</li><li>Initial risks list</li></ul> |
| Second workshop | <ul><li>Additional risks identified</li><li>All risks then categorised</li><li>Connections identified (weak-medium-strong) between the categories</li><li>Risks characterised as strategic or operational</li></ul> | <ul><li>Risks mapped to risk categories</li><li>Connection between the different risk categories</li></ul> |
| Third workshop | <ul><li>Case-based role playing to apply risks and categories to a specific scenario.</li></ul> | |

*Table 2. Workshop activities*

In *Workshop 1* participants identified design elements for a Sourcing Risk and Control Framework, e.g. as QFD has a product development orientation, participants were asked about the type of risk management characteristics important in the development of the framework. Participants then classified these requirements into categories and prioritised them. These were written onto PostIt notes. One category that was identified and formed a substantial part of the workshop discussion was "Defining categories for sourcing risks." All sourcing risks identified throughout the discussion were also written onto PostIt notes (one per note).

In *Workshop 2* participants identified sourcing risks (on additional PostIt notes) building on the work from the previous workshop. These risks were then categorised into 16 high-level categories (see Appendix One). Relationships were then assigned between risk categories i.e. weak to strong relationships and characterised as strategic or operational. Analysis of the PostIt notes also highlighted risks that were more relevant to outsourcing vendors, rather than client organisations.

In *Workshop 3* the risk list and classification was reviewed by a small team of academics from different disciplines and practitioners through its application to a case study scenario. Workshop conversations and PostIT notes were transcribed into a spread-sheet format.

## 4   Findings, discussion and implications

The 16 categories of risks and 151 risk types (see Appendix One) identified in the workshops show that strategy related risks (91) are more prevalent than operational type risks (60). Comparison of these findings to the work of Gandhi et al (2012) and de Sá-Soares et al. (2014) are summarised in Tables 3 and 4 respectively. We compare our findings to these two studies as they provide the most recent comprehensive reviews of the literature.

| Risk category | Schedule (1) | Technical (2) | Financial (3) | Vendor (4) | Culture (5) | Reputation (6) | Intellectual property (7) | Flexibility (8) | Compliance (9) | Quality (10) |
|---|---|---|---|---|---|---|---|---|---|---|
| Strategy | | √ | √ | √ | | √ | √ | √ | √ | |
| Reputation | | | | √ | √ | √√ | | | | |
| Design | | | | | | | | | | |
| Vendor | | | | √ | | | | | | |
| IP | | | | | | | | | | |
| SLA | √ | √ | | √ | | √ | | | | √ |
| Staff | | | | | | | √ | | | |
| Practice | | √ | | | √ | | | | | √ |
| Disaster recovery | | | | √ | | √ | | √ | | |
| ROI | | √ | | | | | | | | |
| Requirements | | √ | | | | | | | | |
| Selection | | √ | | √ | | | | | | |
| Cost risks | √ | | √ | | | | | | | |
| Contract | | | | √ | | | | √ | √ | √ |
| Transition | | | | | √ | | | √ | | |
| Psychological | | | | | | | | | | |

*Table 3. Comparison of risk categories with Gandhi et al. (2012) – client perspective*

In Table 3 we cross reference the 10 risk categories from Gandhi et al.'s (2012) study to the detailed table in Appendix One. Gandhi et al. (2012) did not survey vendor organisations, therefore, as shown in Appendix One (highlighted in bold), we did not map categories related to vendor perceptions. In addition, we shade the risk types that we could not match to Gandhi et al.'s (2012) category descriptions. Some of these unmatched areas could be explained by approaches taken to characterising the risks. For example, we have adopted the labels and classifications used by the participants in our focus group. Gandhi et al. (2012) identified the risks based on a literature review and then conducted a survey to prioritise the risks. Whilst these are important considerations, of particular interest in our findings is that the unmatched areas tend to represent governance related matters, such as accountability issues, control and assurance, strategic alignment and top management support. Further, 60% of the risks identified were characterised as strategic. This is in contrast to Gandhi et al.'s (2012), findings where more than 50% of the risk classifications were operational. These findings appear to point to Lacity et al.'s (2009, 142) 'glimpse' of the future, when they anticipated that a shift from management to leadership would be required if "governance, control, flexibility and superior business outcomes are to be the consequences" of increasing "globalizing and technologizing of the supply of business services." The governance focus is also supported in de Sá-Soares et al. (2014) study of risks of client organisations.

In addition, all but two of the risk categories (design and psychological) identified in our study could be mapped to Gandhi et al.'s (2012) categories. That is, at least one risk type within a category could be identified in Gandhi et al.'s (2012) risk category descriptions. The emotional type of risk represented in the psychological category appears to represent a common view expressed in the literature about the overwhelming number of potential sourcing risks. Lacity et al. (2009, 135) state that practitioners may find the "best way to mitigate risk is through experience." Current theory however tends to emphasise risk and its assessment via normative rules and probabilities, providing limited insight into this experiential view. Theories that represent risk as experiences and emotions (e.g. Slovic et al. 2004; Lupton 2013) as well as consider multiple institutional influences (e.g Thornton et al. 2012) may provide better insights into the dynamics of risk and control related practices within and across organisations and help fuse different approaches to manage sourcing risk more creatively.

Gandhi et al. (2012, 61) found that risks were "somewhat individual" with a limited "extent of overlapping." However, Gandhi et al. (2012, 63) did acknowledge possible interrelationships. For example, financial and reputation risks were identified as having a possible effect on vendor risk. Our results (as shown in Appendix One) indicate that all strategy related risks (both client and vendor) are strongly related to Return On Investment (ROI) and reputational damage and as the levels of strategy related risks increase, the possibility of negative ROI and reputational damage increases. The second highest number of risks (27) relate to contracts. Out of these, 5 risks relate to the vendor perspective, 9 to the client, and 13 to both vendor and client perspectives. These contract related risks have weak ties to all vendor risks but are strongly related to requirements risks, suggesting that a poorly defined contract may not catastrophically affect the vendor (i.e. put them out of business). Increases in the levels of contract risks indicate a higher chance of the contract being incomplete. The results also indicate that vendor risks are strongly related to Intellectual Property (IP) risks, suggesting that the more unprofessional vendor behaviour becomes, the greater the chance the client's data is at risk. Reputational risks are strongly related to selection risks, which indicates that a poor selection of tools, techniques, processes or vendor, can increase the possibility of reputational damage. Our analysis also reveals that ROI risks are strongly related to cost, so if the level of risks related to ROI increases, it is highly likely that the level of cost related to various tasks undertaken for the ICT sourcing project will increase. Transition risks, however, do not relate to any of the 16 risk categories. SLA risks have been identified as being strongly related to requirements risks so that high levels of incompleteness or vagueness increase the likelihood of SLA related risks. Practice risks have not been identified as being related to any particular category of risks. These findings point towards the need for further work examining the inter-relationships of sourcing risks and provide useful insights for designing risk mitigation strategies. The need for an integrated theory of IT-related risk is not new to the IS field (Markus 2000). However, an integrated view of risk control, and specifically in the context of sourcing, remains problematic (Mathew 2011).

In Table 4 we cross reference nine risk categories from de Sá-Soares et al.'s(2014) study. Whilst the de Sá-Soares et al.'s (2014) study also examines risks from a client perspective, we focus here on the vendor perspective. Further, risk is conceptualised in terms of factors (sources of dangers), negative outcomes and undesirable consequences. For the purpose of this comparison we conflate the factors and danger categories together; acknowledging that this is a limitation of our study. Finally, outsourcing risks are characterised as outsourcing stages compared to the strategy/operational view adopted in Gandhi et al. (2012) and our study. Whilst a lifecycle view was not adopted in the first stage of this work, it will be considered in the next phase in developing mitigation strategies.

| Risk category | Capability (1) | Communication (2) | Customer Structure (3) | Environment/competition (4) | Governance (5) | Requirements (6) | Culture (7) | Contract (8) | Trust (9) |
|---|---|---|---|---|---|---|---|---|---|
| Strategy | | | | √ | | | | | |
| Reputation | | | | | | | √ | | |
| Design | | | | | | | | | |
| Vendor | | | | √ | | | | | |
| IP | | | | | | | | | |
| SLA | √ | | | | | | | | |
| Staff | √ | | | | | | | | |
| Practice | √ | √ | | | √ | | | | |
| Disaster/rec | | | | | | | | | |
| ROI | | | | | | | | | |
| Requirements | | | | | √ | √ | | | |
| Selection | | | | | | | | | |
| Cost risks | | | | | | | | | |
| Contract | | | | | | | | √ | |
| Transition | | | | | | | | | |
| Psychological | | | | | | | | | |

*Table 4.Comparison of risk categories to de Sá-Soares et al. (2014) – vendor perspective*

As shown in Table 4, there are a number of risk categories that do not match to the vendor risk categories identified in the de Sá-Soares et al. (2014) study. Whilst this may be partly explained by the approach taken in characterising the risks, it more importantly points towards the "imbalance between the works that identify IS outsourcing elements related to the customers and those related to providers" (de Sá-Soares et al. 2014, 38). Our study confirms the vendor risks identified in de Sá-Soares et al. (2014) extensive literature review, but also provides additional risk types and categories that may contribute to this dearth in the literature.

## 5   Conclusion and future work

In this paper we report on the first stage of our research project, designed to document how sourcing risk is represented in practice in an Australian context. Our workshop approach provided a very effective meta-learning mechanism that has forged multiple perspectives on risk into a cohesive set of relevant categories. These initial categories were found to be broadly consistent with existing classifications mainly constructed from literature reviews. Of more importance, the risk categories identified in this study also build on these existing classifications and will be used as a basis for the identification of enablers, inhibiters and mechanisms to control outsourcing risks; the next stage of this project. This research has also identified potential areas for further theoretical development in terms of: the emotional and experiential nature of sourcing risk; the need for incorporating a multi-stakeholder perspective; examining governance, assurance and accountability; and the inter-connectedness of sourcing risk. These matters await investigation and we hope stimulate further debate.

# 7  Appendix One – Risk types and categories

| Code | Risk | For | Perspective | Tables 3[1], 4[2] |
|---|---|---|---|---|
| R1 | **Strategy risks** have strong relationship with ROI and reputational damage risks | | | |
| R1.1 | Risk of the wrong strategy | Client | Strategy | |
| R1.2 | Risk of ineffective strategy | Client | Strategy | |
| R1.3 | Risk of PESTEL factors not understood in chosen strategy | Client | Strategy | 8 |
| R1.4 | Risk of PESTEL factors being not integrated into the strategy | Client | Strategy | 8 |
| R1.5 | Risk of currency value fluctuation | Client | Strategy | 3 |
| R1.6 | Risk of exchange rate fluctuation | Client | Strategy | 3 |
| R1.7 | Client's risk of loss of competitive differentiation | Client | Strategy | |
| R1.8 | Vendor's risk of loss of competitive differentiation | **Vendor** | Strategy | **4** |
| R1.9 | Risk of strategy not being supported by stakeholders | Client | Strategy | 6 |
| R1.10 | Clients risk of losing strategic alignment | Client | Strategy | |
| R1.11 | Risk of the lack of executive support | Client | Strategy | |
| R1.12 | Risk of the lack of executive sponsorship | Client | Strategy | |
| R1.13 | Risk of the executive relationship being stagnant | Client | Strategy | |
| R1.14 | Risk of enterprise architecture misalignment with strategy | Client | Strategy | 2 |

---

[1] Categories of risk from Gandhi et al (2012) numbered in Table 3.

[2] Categories of risk from de Sá-Soares et al. (2014) numbered in Table 4

| Code | Risk | For | Perspective | Tables 3¹, 4² |
|---|---|---|---|---|
| R1.15 | Risk of disruptive technology | Client | Strategy | 8 |
| R1.16 | Risk of technology obsolescence | Client | Strategy | 2 |
| R1.17 | Vendor's risk of technology change | **Vendor** | Strategy | |
| R1.18 | Client's risk of technology change | Client | Strategy | 8 |
| R1.19 | Risk of vendor lock-in | Client | Strategy | 4 |
| R1.20 | Risk of complex technology | Client | Strategy | 2 |
| R1.21 | Risk of overall interoperability for business | Client | Strategy | 2 |
| R1.22 | Risk of overall interoperability for IT | Client | Strategy | 2 |
| R1.23 | Risk of untested technology | Client | Strategy | 2 |
| R1.24 | Risk of knowledge retention | Client | Strategy | 7 |
| R1.25 | Client's risk of losing expertise | Client | Strategy | 7 |
| R1.26 | Risk of poor understanding of the risk and reward trade-offs | Client | Strategy | |
| R1.27 | Risk of non-compliance with the law | Client | Strategy | 9 |
| R1.28 | Risk of strategic alliances | Client | Strategy | 4 |
| R1.29 | Risk of moving to a new business model | Client | Strategy | |
| R1.30 | Risk of commercial model (test vs. outcome) | Client | Strategy | |
| R1.31 | Risk of poor understanding of outsourcing strategy risks | Client | Strategy | |
| R1.32 | Risk of poor understanding of commercial acumen | Client | Strategy | |
| R1.33 | Risk of management holding onto business model | Client | Strategy | |
| R1.34 | Risk of outsourcing strategy vs strategy to outsource | Client | Strategy | |
| R2 | **Reputational damage risks** have strong relationship with strategy and selection risks | | | |
| R2.1 | Risk of prospective vendors unethical behaviour | Client | Strategy | 4 |
| R2.2 | Risk of cultural incompatibility | Both | Strategy | 5,7 |
| R2.3 | Risk of reputation loss through outsourcing | Client | Strategy | 6 |
| R2.4 | Risk of client exploiting contract gaps | **Vendor** | Strategy | |
| R2.5 | Risk of reputation damage caused by vendor actions | Client | Strategy | 6 |
| R3 | **Design risks** | | Strategy | |
| R4 | **Vendor risks** are strongly related to IP risks however weakly related contract risks | | | |
| R4.1 | Risk of service provider market concentration | Client | Strategy | 4 |
| R4.2 | Risk of lack of competition | Client | Strategy | 4 |
| R4.3 | Risk of vendor going bankrupt | Client | Strategy | 4 |
| R4.4 | Risk of vendor opportunistic behaviour | Client | Strategy | 4 |
| R4.5 | Risk of moving cost to OPEX | Client | Strategy | 4 |
| R4.6 | Risk of competitors outperforming | Client | Strategy | 4 |
| R4.7 | Risk of regulations | **Vendor** | Strategy | 4 |
| R4.8 | Risk of customer market concentration | Client | Strategy | 4 |
| R4.9 | Risk of vendor monopoly | Client | Strategy | 4 |
| R5 | **IP risks** have strong relationship with vendor and contract risks | | | |
| R5.1 | Risk of withholding information | Client | Operational | 7 |
| R5.2 | Risk of data offshoring | Client | Operational | 7 |
| R6 | **SLA risks** are strongly related to requirements risks | | | |
| R6.1 | Risk of improved service quality | Client | Operational | 10 |
| R6.2 | Risk of service disruptions | Client | Operational | 2 |
| R6.3 | Risk of poor service quality | Client | Operational | 10 |
| R6.4 | Risk of service delivery failure | Client | Operational | 4 |
| R6.5 | Risk of service level agreement failures | Both | Operational | 10 |
| R6.6 | Risk of entire project failing | Both | Operational | |
| R6.7 | Risk of poor relationship between bus. drivers and IT services | Both | Operational | |
| R6.8 | Risk of lack of clarity of vendor governance | Client | Operational | |
| R6.9 | Risk of lack of clarity of vendor management | Client | Operational | |
| R6.10 | Risk of lack of clarity of contract management | Client | Operational | |

| Code | Risk | For | Perspective | Tables 3¹, 4² |
|---|---|---|---|---|
| R6.11 | Risk of lack of clarity of operations management | Client | Operational | |
| R6.12 | Risk of lack of contract management/service delivery | Both | Operational | N/A,1 |
| R6.13 | Risk of lack of accountability between/across vendors | Client | Operational | |
| R6.14 | Risk of lack of internal accountability | Client | Operational | |
| R6.15 | Risk of poor deliverable quality | Client | Operational | 10 |
| R6.16 | Risk of vendor failing to deliver | Client | Operational | 1 |
| R6.17 | Risk of losing private information | Client | Operational | 6 |
| R6.18 | Risk of vendor misusing client data | Client | Operational | 6 |
| R6.19 | Risk of information loss | Client | Operational | 6 |
| R6.20 | Risk of losing confidential information | Both | Operational | 6 |
| R6.21 | Risk of insufficient monitoring | Both | Operational | 10 |
| R6.22 | Risk of insufficient reporting | Both | Operational | 10 |
| R6.23 | Risk of on-demand-capacity | Both | Operational | 2 |
| R6.24 | Risk of loss of operative capacity | Both | Operational | 2 |
| R6.25 | Risk of end to end governance of supplier portfolio vs. individual contracts | Both | Operational | |
| R7 | **Staff risks** | | | |
| R7.1 | Risk of key personnel missing | Client | Operational | |
| R7.2 | Risk of key personnel leaving to work for the vendor | Client | Operational | 7 |
| R7.3 | Risk of key personnel leaving to work for the competitor | Both | Operational | 7,1 |
| R7.4 | Risk of specialist skills residing with the customer | **Vendor** | Operational | 1 |
| R7.5 | Risk of access to specialist expertise | Client | Operational | 7 |
| R7.6 | Risk of key personnel leaving to work for the client | Vendor | Operational | |
| R8 | **Practice risks** | | | |
| R8.1 | Risk of mismatching working practices | Both | Operational | N/A,1 |
| R8.2 | Risk of mismatching delivery methodology | Both | Operational | N/A,1 |
| R8.3 | Risk of ineffective delivery methodology | Client | Operational | 10 |
| R8.4 | Risk of unreliable measurement | Client | Operational | 10 |
| R8.5 | Risk of culture differences | Both | Operational | 5 |
| R8.6 | Risk of organisational culture differences | Both | Operational | 5 |
| R8.7 | Risk of miscommunication due to time zone differences | Both | Operational | 5,2 |
| R8.8 | Risk of miscommunication due to geographical distances | Both | Operational | 5,2 |
| R8.9 | Risk of loss of functionality | Client | Operational | 2 |
| R8.10 | Risk of loss of control over IT operations | Client | Operational | |
| R8.11 | Risk of work practices misalignment | Both | Operational | 5,5 |
| R8.12 | Risk of improving time to market (positive) | Client | Operational | |
| R8.13 | Risk of assurance | Client | Operational | |
| R8.14 | Risk of retained capability (positive) | Client | Operational | |
| R9 | **Disaster recovery risks** | | | |
| R9.1 | Risks of natural disasters | Both | Operational | 8 |
| R9.2 | Risks of loss of data traceability in case of disaster | Both | Operational | 6 |
| R9.3 | Risk of cross regional issues related to disaster recovery | Both | Operational | 4 |
| R9.4 | Risk of business continuity in case of disaster | Both | Operational | 4 |
| R10 | **ROI** risks are strongly related to cost and strategy risks | | | |
| R10.1 | Risk of agreeing on Pay-as-you-go (positive) | Client | Strategy | |
| R10.2 | Risk of gaining better ROI (positive) | Client | Strategy | 3 |
| R10.3 | Risk of improved uptime (positive) | Client | Strategy | |
| R10.4 | Risks of unexpected (negative) financial outcomes | Client | Strategy | 3 |
| R11 | **Requirements risks** are strongly related to SLA, contract and selection risks | | | |
| R11.1 | Risk of failure to provide access to suitable resources | Both | Operational | 2 |
| R11.2 | Risk of complex concepts resulting in misunderstanding | Both | Operational | |
| R11.3 | Risk of poorly understood requirements | Both | Operational | N/A,6 |
| R11.4 | Risk of changing requirements | Both | Operational | |

| Code | Risk | For | Perspective | Tables 3[1], 4[2] |
|---|---|---|---|---|
| R12 | **Selection risks** are strongly related to **reputational damage** risks | | | |
| R12.1 | Risk of selector bias | Both | Strategy | 4 |
| R12.2 | Risk of selection of wrong tools | Both | Strategy | 2 |
| R12.3 | Risk of selection of wrong configuration of systems | Both | Strategy | 2 |
| R12.4 | Risk of power differences between vendor and client | Both | Strategy | 4 |
| R12.5 | Risk of adverse selection | Client | Strategy | 4 |
| R12.6 | Risk of unfair selection process | **Vendor** | Strategy | |
| R12.7 | Risk of uneducated client | **Vendor** | Strategy | |
| R12.8 | Risk of unclear scope during selection process | Both | Strategy | 4 |
| R12.9 | Risk of lack of defined roles and responsibilities in sourcing | Client | Strategy | 4 |
| R13 | **Cost risks** are strongly related to ROI risks | | | |
| R13.1 | Risk of cost overruns | Both | Operational | 3 |
| R13.2 | Risk of poor estimation | Both | Operational | 3 |
| R13.3 | Risk of delays | Both | Operational | 1 |
| R13.4 | Risk of hidden costs | Both | Operational | 3 |
| R14 | **Contract risks** are strongly related to requirements risks but weakly related to vendor risks | | | |
| R14.1 | Risk of poorly formed SLA's leading to contract confusion | **Vendor** | **Strategy** | 8 |
| R14.2 | Risk of poorly considered legal framework | Both | Strategy | 10 |
| R14.3 | Risk of misunderstanding the contract | Both | Strategy | 10 |
| R14.4 | Risk of no mechanisms to protect against failure | Client | Strategy | |
| R14.5 | Risk of time to deliver | **Vendor** | Strategy | |
| R14.6 | Risk of contracting and sub-contracting | Client | Strategy | |
| R14.7 | Risk of law breach leading to prosecution | Client | Strategy | 10 |
| R14.8 | Risk of undefined requirements or needs from client | **Vendor** | Strategy | |
| R14.9 | Risk of multiple vendors | Client | Strategy | 4 |
| R14.10 | Risk of multiple vendors blaming each other for failures | Client | Strategy | 4 |
| R14.11 | Risk of contract complexity (too many and too varied) | Both | Strategy | 10 |
| R14.12 | Risk of early termination penalties | Vendor | Strategy | |
| R14.13 | Risk of vendor bankruptcy or takeover | Client | Strategy | 4 |
| R14.14 | Risk of contract lock-in | Client | Strategy | 8 |
| R14.15 | Risk of false sense of risks being mitigated or transferred | Both | Strategy | |
| R14.16 | Risk of M&A activity impacting client/service provider strategy | Both | Strategy | |
| R14.17 | Risk of scalability (positive) | Client | Strategy | 8 |
| R14.18 | Risk of incomplete contracts | Both | Strategy | |
| R14.19 | Risk of APS 231 | Client | Strategy | |
| R14.20 | Risk of contract deficiencies | Both | Strategy | |
| R14.21 | Risk of undefined measurements | Both | Strategy | 10 |
| R14.22 | Risk of customer bankruptcy | **Vendor** | Strategy | |
| R14.23 | Risk of local and international regulations | Both | Strategy | 9 |
| R14.24 | Risk of conflicts of law | Both | Strategy | 9 |
| R14.25 | Risk related to jurisdiction | Both | Strategy | 9 |
| R14.26 | Risk of identity | Both | Strategy | |
| R14.27 | Risk of accountability and responsibility being undefined | Both | Strategy | |
| R15 | **Transition risks** | | | |
| R15.1 | Risk of customer decision to insource | **Vendor** | Strategy | |
| R15.2 | Risk of loss of employee morale during transition | Both | Strategy | 5 |
| R15.3 | Risk of reverse transition | Client | Strategy | 8 |
| R16 | **Psychological risks** | | | |
| R16.1 | Risk of risk professionals having a nervous breakdown | Both | Operational | |